\documentclass{article}  

\usepackage{graphicx}
\usepackage{amsthm}
\usepackage{amsmath,amssymb}
\usepackage{graphicx}
\usepackage{subcaption}
\usepackage{enumitem} 
\usepackage{mathtools}
\usepackage{enumitem}
\usepackage{algorithm}
\usepackage{textcomp}
\usepackage{siunitx}

\newtheorem{Definition}{Definition}

\newcommand{\LQRTrack}[1]{u_\mathrm{LQR}^{#1}}

\newcommand{\Sol}[2]{\Sigma(#1, #2)}

\newcommand\norm[1]{\left\lVert#1\right\rVert}

\begin{document}

\title{LQR-trees with Sampling Based Exploration of the State Space \thanks{This work was supported by the project GA21-09458S of the Czech Science Foundation GA \v{C}R and institutional support
RVO:67985807.}}

\author{Ji{\v r}{\'\i} Fejlek$^{1,2}$\footnote{ORCID: 0000-0002-9498-3460} \;and Stefan Ratschan$^{1}$\footnote{ORCID: 0000-0003-1710-1513}\\ 
$^1$The Czech Academy of Sciences, Institute of Computer Science,\\ Pod Vod\'{a}renskou v\v{e}\v{z}\'{i} 271/2, Prague, 182 07, Czech Republic\\
$^2$Czech Technical University in Prague, Faculty of Nuclear Sciences and Physical Engineering, \\Department of Mathematics, Trojanova 13, Prague, 120 00, Czech Republic}

\maketitle
\thispagestyle{empty}
\pagestyle{empty}


\begin{abstract}
This paper introduces an extension of the LQR-tree algorithm, which is a feedback-motion-planning algorithm for stabilizing a system of ordinary differential equations from a bounded set of initial conditions to a goal. The constructed policies are represented by a tree of exemplary system trajectories, so called demonstrations, and linear-quadratic regulator (LQR) feedback controllers. Consequently, the crucial component of any LQR-tree algorithm is a demonstrator that provides suitable demonstrations. In previous work, such a demonstrator was given by a local trajectory optimizer. However, these require appropriate initial guesses of solutions to provide valid results, which was pointed out, but largely unresolved in previous implementations. In this paper, we augment the LQR-tree algorithm with a randomized motion-planning procedure to discover new valid demonstration candidates to initialize the demonstrator in parts of state space not yet covered by the LQR-tree. In comparison to the previous versions of the LQR-tree algorithm, the resulting exploring LQR-tree algorithm reliably synthesizes feedback control laws for a far more general set of problems.
\end{abstract}


\section{Introduction}

This paper provides an algorithm that automatically synthesizes a feedback control policy for systems of ordinary differential equations that stabilizes the system into the goal. The algorithm is based on the construction of LQR-trees~\cite{LQRtrees1,LQRtrees2} that consist of a tree of trajectories, so called demonstrations~\cite{Rav:19}, stabilized using linear quadratic regulator (LQR) feedback control laws. These stabilized demonstrations are used to cover the investigated set: Either a conservative region of attraction of the goal set can be computed for each demonstration~\cite{LQRtrees1}, or these regions can be estimated via system simulations~\cite{LQRtrees2}. 

The key component in both versions of this algorithm is a demonstrator, a procedure that provides control inputs that steer the system into the goal set. In previous work~\cite{LQRtrees1,LQRtrees2,Fejlek:22}, this demonstrator took the form of a trajectory optimization method~\cite{Bet:10}. However, such solvers require appropriate initial guesses to provide valid results. If these guesses are not good enough, the demonstrator regularly fails, causing the LQR-tree to progress slowly, if at all. This was the case in the implementation of LQR-trees in~\cite{LQRtrees2}, where the demonstrator was initialized with failed system simulations generated during the LQR-tree algorithm. This turned out not to be sufficient~\cite{LQRtrees2} and our computational experiments support this observation.

In this paper, we remedy this issue by extending the LQR-tree algorithm with exploration of the state space based on randomized motion-planning. More specifically, we use  rapidly-exploring random trees (RRT)~\cite{RRT4} to further extend the LQR-tree into areas of the state space that are not yet stabilized by the current LQR-tree. The newly discovered connections then serve as initial solutions to the original demonstrator based on a trajectory optimization method. This demonstrator subsequently provides finalized optimized demonstrations that are added to the LQR-tree.

We provide computational experiments on several examples of dimension up to twelve that illustrate the practical applicability of the method and we compare our generation of demonstrations to the one in~\cite{LQRtrees2}. In this comparison, the exploring LQR-tree algorithm is more reliable and often significantly faster.

The structure of the~paper is as follows. In~Section~\ref{sec:problem}, we state the precise problem and discuss related work. In~Section \ref{sec:LQR}, we briefly describe the LQR-tree algorithm. In Section~\ref{sec:RRT}, we present the main contribution of this paper, the integration of randomized motion-planning (RRT) into the generation of demonstrations for the LQR-tree algorithm. In Section~\ref{sec:exp}, we describe the details of our implementation and provide computational experiments. And finally, Section~\ref{sec:conclusion} contains a~conclusion. 

\section{Problem Statement and Related Work}
\label{sec:problem}
Assume a~control system  
\begin{equation}
\label{system}
\dot{x} = F(x,u) 
\end{equation}
where $F\colon\mathbb{R}^n\times \mathbb{R}^m\mapsto\mathbb{R}^n$ is a~smooth function. Let $U \subset \mathbb{R}^m$ be a~bounded set of control inputs. For a~given initial point $x_0$, a~time horizon $T>0$ and a~continuous control input $u\colon[0, T]\mapsto U$, we denote the corresponding solution of~(\ref{system}) by $\Sol{x_0}{u}$. We use the~same notation for the~solution with a~feedback control law $u\colon \mathbb{R}^n\times[0, T]\mapsto U$.

Let $I\subset \mathbb{R}^n$ be a compact set of initial states, $O\subset \mathbb{R}^n$ be a set of obstacles and $G \subset \mathbb{R}^n$ be a~compact set of goal states. Our goal is to construct a~feedback control law $u\colon \mathbb{R}^n\times[0, T]\mapsto U$ such that $\Sol{x_0}{u}$ reaches $G$ for all $x_0 \in I\setminus G$ while staying outside of $O$. 

We base our construction on LQR-trees~\cite{LQRtrees2,LQRtrees1,Fejlek:22}. These use a library of demonstrations consisting of system trajectories with desired control inputs from various states in $I$ and generalize these demonstrations to all states in $I$ using a tracking based on linear-quadratic regulator (LQR)~\cite{Lib:11}.

The construction of an LQR-tree requires a demonstrator~\cite{Rav:19} that generates suitable demonstrations that reach the goal set $G$ under all given constraints. This demonstrator is usually based on trajectory optimization~\cite{Bet:10} that reduces the problem to nonlinear programming (NLP). However, such solvers require appropriate initial guesses, for example,  taken from previous system simulations generated during the LQR-tree construction~\cite{LQRtrees2}. However, such heuristics often fail~\cite{LQRtrees2}, an observation that the computational experiments in this paper confirm. 
This causes the construction of the LQR-tree to slow down, making it difficult for the LQR-tree to cover the whole set $I$. Hence, as stated in~\cite{LQRtrees2}, ``... a motion planner ... robust to poor initialization is therefore desirable.''

Hence we propose to exploit sampling based path planning algorithms~\cite{RRT5}, namely rapidly-exploring random trees (RRT)~\cite{RRT4}, to further explore the state space and find new demonstrations during the LQR-tree construction. Rapidly-exploring random trees are known to efficiently find solutions to path-planning problems both in theory~\cite{RRT12} and in practice~\cite{RRT13}.

\section{LQR-trees}
\label{sec:LQR}
In this section, we describe the LQR-tree algorithm~\cite{LQRtrees1,LQRtrees2,Fejlek:22} in more details, introducing the necessary background and more detailed motivation for our contribution, which we will introduce in the next section. We reduce our attention to the simulation-based variants~\cite{LQRtrees2,Fejlek:22}, but our contribution applies to~\cite{LQRtrees1} as well. The LQR-tree algorithm constructs an LQR-tree, which consists of a set of demonstrations and a set of LQR tracking controllers to follow these demonstrations. Thus, the LQR-tree provides a feedback control law for any state in $I$ that chooses a suitable demonstration and applies the corresponding feedback tracking controller. Provided that the LQR-tree is well designed, the resulting feedback has the desired property, reaching $G$ from $I$ without visiting $O$. 

Let us first specify the form of demonstrations that an LQR-tree consists of. They are open loop trajectories as defined in~\cite{LQRtrees1,Fejlek:22}, extended here to also include obstacles.
\begin{Definition}
\label{demonstration_def}
Let $X \subset \mathbb{R}^n$ and $U \subset \mathbb{R}^m$ be bounded sets, let  $O\subset \mathbb{R}^n$ be a set of obstacles, and let $T > 0$. A~\emph{demonstration from $x_0 \in X\setminus G$} is a~system trajectory $(x,u)$ with $x\colon [0, T] \mapsto X$ and $u\colon [0, T]\mapsto U$, both continuous, that~satisfies the control system~(\ref{system}), and for which $x(0)=x_0$, $x(T)\in G$, and $x(t)\notin O$ for all $t\in [0, T]$.  
\end{Definition}

Assume a set of demonstrations $\mathcal{T}$. For each demonstration $(\tilde{x},\tilde{u})\in \mathcal{T}$, we consider a tracking controller based on LQR~\cite{Lib:11}. We denote the LQR tracking controller by~$\LQRTrack{(\tilde{x},\tilde{u})}$. This LQR controller is characterized by cost matrices $Q, Q_0 \in \mathbb{R}^{n\times n}$ and $R \in \mathbb{R}^{m\times m}$ that we will assume to be positive definite and the same for all demonstrations. We denote the corresponding value function assigning to each state its cost-to-go by $V_{(\tilde{x},\tilde{u})}$.

Let $x_0 \in I$ be a state from which we wish to reach the goal states using the demonstrations in $\mathcal{T}$. In the presence of multiple demonstrations, we need to choose a specific one to follow which we refer to as the \emph{target demonstration}. To that end we use a definition~\cite{Fejlek:22} that generalizes more concrete rules used earlier~\cite{LQRtrees1,LQRtrees2}. 
\begin{Definition}
\label{assignmentrule}
An \emph{assignment rule} $\phi$ is a~function that~for any set of demonstrations $\mathcal{T}$ and point $x\in \mathbb{R}^n$ selects an~element from $\mathcal{T}$. 
\end{Definition}
The LQR value functions $V_{(\tilde{x},\tilde{u})}$ provide a suitable basis for such an assignment rule~\cite{LQRtrees1}. We define
\begin{equation}
\label{LQRarule}
\phi_\mathrm{LQR}(x,\mathcal{T}) = \underset{(\tilde{x},\tilde{u})\in\mathcal{T}}{\mathrm{arg\,min}} V_{(\tilde{x},\tilde{u})}(x)
\end{equation}

The triplet of the set of demonstrations, the corresponding set of LQR tracking controllers and the assignment rule $\phi$ defines a feedback control law for all states in $I$. More specifically, the feedback control law  to steer the system from $x_0\in I$ is given for all $x\in\mathbb{R}^n$ and all $t \in [0, T]$ by $u(x,t) = \LQRTrack{\phi(x_0)}(x,t)$. We further refer to the triplet as an LQR-tree.

Assume an initial set of demonstrations $\mathcal{T}$ which is used to construct an LQR-tree. In general, such an LQR-tree does not provide a feedback control law that steers all states in $I$ to the goal set $G$ while avoiding obstacles $O$. To determine which demonstrations must be further added to $\mathcal{T}$, we consider an algorithm based on sampling of the initial set $I$~\cite{LQRtrees2,Fejlek:22}. 

Assume a sample $x_0$. The algorithms tests via a simulation of the system, whether the feedback control law given by a set of demonstrations $\mathcal{T}$ reaches the goal set $G$ from $x_0$ while staying outside of $O$, which we refer further as the LQR-tree \emph{being successful for} $x_0$. If the LQR-tree is not successful for $x_0$, the algorithm uses the demonstrator to generate a new demonstration from $x_0$, see Algorithm~\ref{alg:LQR-tree}. 

\begin{algorithm}[htb]
\caption{LQR-tree algorithm}
\label{alg:LQR-tree}
\begin{description}
\item[In:] A~control system $\dot{x} = F(x,u)$, assignment rule $\phi$, cost matrices $Q, Q_0 \in \mathbb{R}^{n\times n}$ and $R \in \mathbb{R}^{m\times m}$ 
\item[Out:] A~feedback control law in the form of an LQR-tree given by a set of demonstrations $\mathcal{T}$
\end{description}
\begin{enumerate}
\item Generate an initial set of demonstrations $\mathcal{T}$ and construct an initial LQR-tree
\item Choose an open bounded set $W \supset I$ and a sequence $M$ from $W$ whose set of elements is dense in $W$
\item Iterate
\begin{enumerate}
\item Let $x_0$ be the next sample from the sequence $M$. Check via a simulation of the system, if the LQR-tree given by $\mathcal{T}$ is successful for $x_0$
\item If the LQR-tree given by $\mathcal{T}$ is not successful for $x_0$, generate a new demonstration from the counterexample $x_0$ and add it to $\mathcal{T}$
\end{enumerate}
\end{enumerate}
\end{algorithm}

Assume that all demonstrations have a certain leeway, a \emph{clearance}~\cite{RRT12}, for tracking controllers to meet all the constraints while tracking target demonstrations~\cite{LQRtrees1}. More precisely, assume that for some fixed $\delta >0$, all demonstrations $(\tilde{x},\tilde{u})$ meet $B_{\tilde{x}(t)}^\delta \notin O, B_{\tilde{u}(t)}^\delta \in U$ for all $t\in[0, T]$, and $B_{\tilde{x}(T)}^\delta \in G $, where $B^\delta_x$ denotes an open ball with a centre in $x$ and a diameter $\delta$. Then under some mild assumptions on the assignment rule $\phi$~\cite{Fejlek:22}, Algorithm~\ref{alg:LQR-tree} produces a feedback control law that reaches the goal set $G$ while avoiding obstacles $O$ after finitely many iterations~\cite{Fejlek:22}, that is, after adding finitely many demonstrations to $\mathcal{T}$.

A key part of successful implementation of Algorithm~\ref{alg:LQR-tree} is a demonstrator that provides new demonstrations for the LQR-tree. The implementation~\cite{LQRtrees2} uses a demonstrator based on trajectory optimization~\cite{Bet:10} that is initialized from the failed simulations obtained in step 3a. However, as observed in~\cite{LQRtrees2} for a pendulum on cart example, only about 30\% of all attempts of the demonstrator succeeded which indicates that many initializations were poor. 

\begin{figure}
\centering
\includegraphics[width=0.75\textwidth]{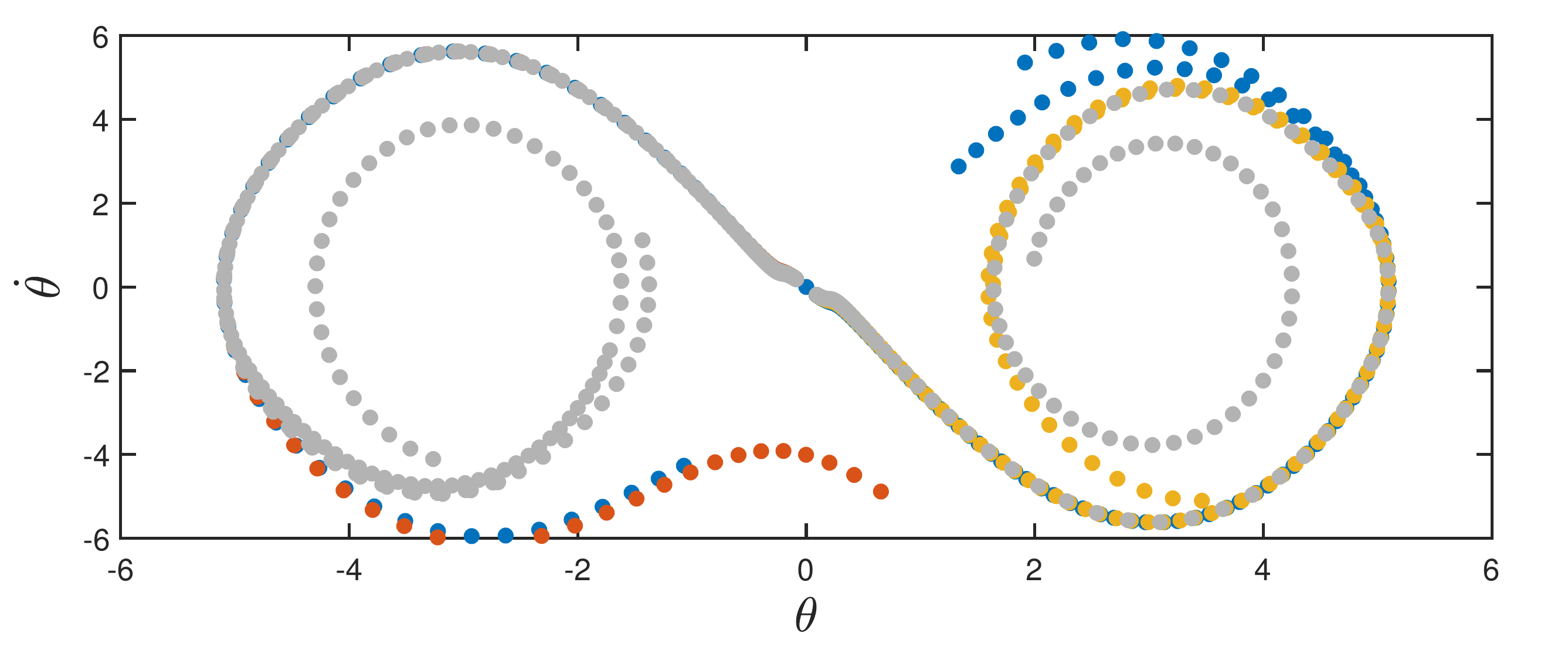}
\caption{LQR-tree for the pendulum after 100 (blue), 200 (red), 300 (yellow) and 400 (gray) calls of the demonstrator.}
\label{fig:slow_progress}
\end{figure}

To demonstrate this fact further, assume the following inverted pendulum example
\begin{equation}
\label{inverted_pendulum}
\ddot{\theta} = \frac{1}{ml^2}\left( u + mgl\sin\theta -b\dot{\theta}\right),
\end{equation}
where $m = 0.5$, $l = 1$, $b = 0.1$ and $g = 9.81$. In this example, we wish to steer the system to $[0,0]$ from the initial set $I = [-4,4] \times [-5,5]$ with the control inputs bounded by $[-1.25, 1.25]$, which we restrict further for demonstrations to $[-1, 1]$. We set the LQR costs as $Q = Q_0 = I$ and $R = I$ and start the LQR-tree algorithm with a demonstration $\tilde{x} = 0$ and $\tilde{u} = 0$. We base the demonstrator on a direct optimal control solver~\cite{Bet:10} with the same quadratic cost. The NLP problem is constructed via the Hermite-Simpson collocation method~\cite{Har:87}. We set the length of demonstrations as $T = 10$ with a time step of $h = 0.05$. We initialize the NLP solver via failed simulations. The failure rate of the demonstrator is very high (97\%) and hence, the LQR-tree algorithm struggles to make progress, as can be seen in Fig.~\ref{fig:slow_progress}.

\section{Exploring LQR-trees}
\label{sec:RRT}
To solve the issue of generating new demonstrations in step 3b of the LQR-tree algorithm, which is crucial for its efficiency and convergence, we will exploit rapidly-exploring random trees (RRT)~\cite{RRT4}. The RRT algorithm explores the state-space and generates valid initial estimates of the demonstrations which we pass on to the NLP demonstrator whose result then forms new demonstrations.

The RRT algorithm is an incremental algorithm that creates a tree in the state space by connecting new states to already explored ones. Its main strength is that the RRT algorithm does not require a steering function~\cite{RRT5}, a function that connect two states with a dynamically feasible trajectory, which generally requires a solution to a boundary value problem. Instead, the RRT algorithm extends the tree by propagating the system via chosen control inputs~\cite{RRT12,RRT15}. 

\begin{figure}
\centering
\includegraphics[width=0.75\textwidth]{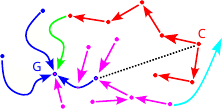}
\caption{Sketch of Algorithm~\ref{LQR-tree gen}}
\label{LQRRRT_fig}
\end{figure}

Let us assume that Algorithm~\ref{alg:LQR-tree}, after some iterations, has arrived at the set of demonstrations $\mathcal{T}$ corresponding to the blue trajectories in Fig.~\ref{LQRRRT_fig}, all terminating in the goal (G). Assume that we now need to generate a new demonstration from a counterexample (C) whose target demonstration is indicated by the end of the black dotted line. For this, we need to properly initialize the NLP based demonstrator. Also, we want to  avoid calling the demonstrator too often, since it is costly.

We now use the RRT algorithm to explore the state space by building two trees: a counterexample tree (red) that is built from the counterexample and a demonstration tree (magenta) that is built from a discrete sampling of one of demonstrations in $\mathcal{T}$---initially the given target demonstration. 

During the growth of both trees, for each newly visited state, we test whether a system simulation using the current LQR-tree results in a new demonstration from this state. In the case of the counterexample tree, we search for states for which the current LQR-tree is successful  (green). Analogously, in the case the demonstration tree, we search for states from which the current LQR-tree feedback control law fails (cyan). In the first case, we obtain a valid system trajectory from a counterexample to the goal. The second case expands the LQR-tree into new areas where the original feedback control law failed and thus, it further enlarges the parts of the state space where the first case can occur. In both cases, the recovered system trajectories serve as an initial solution for the NLP based demonstrator, which provides the final demonstrations that are added to the LQR-tree. We further denote calls to the demonstrator by a function $demo$. Note that demonstrations must have a clearance to assure that the LQR-tree is successful for all states that are close enough to the demonstrations, and thus the simulation must be tested with this clearance in mind.

The resulting algorithm that explores the state space and generates new demonstrations in step 3b of the LQR-tree algorithm is Algorithm~\ref{LQR-tree gen}.

\begin{algorithm}[htb]
\caption{Exploring LQR-trees (step 3b of Algorithm~\ref{alg:LQR-tree})}
\label{LQR-tree gen}
\begin{description}
\item[In:] $\mathcal{T}$ without a demonstration from $x_\mathrm{counter}$
\item[Out:] $\mathcal{T}$ with a demonstration from $x_\mathrm{counter}$
\end{description}
\begin{enumerate}
\item View  each demonstration in $\mathcal{T}$ and the counter-example $x_\mathrm{counter}$ as corresponding RRT trees 
\item Let $\tilde{x}$ be the first element of the pair $\phi_\mathrm{LQR}(x_\mathrm{counter}, \mathcal{T})$
\item Let $\mathcal{R}_{\mathrm{counter}}$ be the RRT tree containing $x_\mathrm{counter}$
\item Let $\mathcal{R}_{\mathrm{dem}}$ be the RRT tree containing $\tilde{x}$
\item Repeat until $\mathcal{T}$ contains a demonstration from $x_\mathrm{counter}$
\begin{enumerate}
\item Bidirectionally expand $\mathcal{R}_{\mathrm{dem}}$ and  $\mathcal{R}_{\mathrm{counter}}$
\item During expansion of the RRT trees, for every added node $x$
\begin{enumerate}
\item If $x$ was added to $\mathcal{R}_{\mathrm{counter}}$ and the LQR-tree  is successful for $x$ with a clearance
\begin{itemize}
\item[] Add $demo(ps)$ to $\mathcal{T}$, where $p$ is the path from $x_\mathrm{counter}$ to $x$ in $\mathcal{R}_{\mathrm{counter}}$ and $s$ is the simulation from $x$ to $G$ using the LQR-tree
\end{itemize}
\item If $x$ was added to  $\mathcal{R}_{\mathrm{dem}}$ and the LQR-tree  is \emph{not} successful for $x$  
\begin{itemize}
\item[] Add $demo(p\tilde{x})$ to $\mathcal{T}$, where $p$ is the path from $x$ to $\tilde{x}(0)$ in $\mathcal{R}_{\mathrm{dem}}$, also viewing the added demonstration as another RRT tree.
\end{itemize}
\end{enumerate}
\item Update $\tilde{x}$ to the first element of\begin{equation*}
\underset{(\tilde{x},\tilde{u})\in\mathcal{T}}{\mathrm{arg\,min}}
\underset{\;\vphantom{(} x\in \mathcal{R}_{\mathrm{counter}}\hspace*{-0.3cm}}{\mathrm{min}\vphantom{g}}
V_{(\tilde{x},\tilde{u})}(x)
\end{equation*}
 and $\mathcal{R}_{\mathrm{dem}}$ to the LQR tree containing $\tilde{x}$.
\end{enumerate}
\end{enumerate}
\end{algorithm}

Concerning the concrete variant of the RRT algorithm to use here, there are numerous possibilities that extend the original RRT algorithm with the aim of improving its performance~\cite{RRT6,RRT7}. We need a variant that works well for general kinodynamic problems. We also want to exploit the presence of another RRT-like tree, the current LQR-tree, to grow it backwards, trying to connect it to the original RRT tree. 

For this, we adapt bidirectional RRT-connect~\cite{RRT16} which is a popular~\cite{RRT18} greedy variant of RRT. In this algorithm, the first tree, which we refer to as primary is extended in a random direction $x_\mathrm{rand}$ (generated uniformly from $X$) using the so called best control inputs~\cite{RRT5}. We use a discretely sampled set of actions, searching for the best action via systematic enumeration.

If the node fails to extend closer to $x_\mathrm{rand}$, the whole process is repeated. If the node extends successfully, the algorithm continues to extend the tree until it fails. Afterwards, the secondary tree is extended in the direction of the last extension of the primary tree. Finally, the primary tree and secondary are swapped and the process repeats. See Algorithm~\ref{LQRRRT} for the full description.

\begin{algorithm}[htb]
\caption{bidirectional RRT-connect}
\label{LQRRRT}
\begin{description}
\item[In:] RRT trees $\mathcal{R}_\mathrm{counter}$, $\mathcal{R}_\mathrm{dem}$
\item[Out:] Extended trees $\mathcal{R}_\mathrm{counter}$ and $\mathcal{R}_\mathrm{dem}$
\item[Param:] Metric $\eta$, state space $X$ and set of control inputs~$U$
\end{description}

\begin{enumerate}
\item Let $\mathcal{R}_\mathrm{prim}=\mathcal{R}_\mathrm{counter}$ (the primary tree) and $\mathcal{R}_\mathrm{sec}= \mathcal{R}_\mathrm{dem}$ (the secondary tree)
\item Repeat for a given number of iterations
\begin{enumerate}
\item Repeat
\begin{itemize}
\item[]  Grow the primary tree $\mathcal{R}_\mathrm{prim}$ from  $x_\mathrm{near}$ to $x_\mathrm{rand}$, where $x_\mathrm{rand}$ is a random state in $X$ and $x_\mathrm{near}$ is the nearest state to $x_\mathrm{rand}$ wrt. $\eta$ in the tree $\mathcal{R}_\mathrm{prim}$ 
\end{itemize}
until this results in an expansion of $\mathcal{R}_\mathrm{prim}$ 
\item Grow the secondary tree $\mathcal{R}_\mathrm{sec}$ from $y_\mathrm{near}$ to $x$, where $x$ is the node resulting from the last expansion of the primary tree and $y_\mathrm{near}$ is the nearest state to $x$ wrt. $\eta$ in the tree $\mathcal{R}_\mathrm{sec}$
\item Swap the primary and the secondary tree
\end{enumerate}
\end{enumerate}

\vspace*{0.3cm}
\textbf{Sub-algorithm}: Growing the tree $\mathcal{R}$ from $p\in\mathcal{R}$ to $q\not\in\mathcal{R}$:\\[2mm]
Loop
\begin{enumerate}
\item Let $u = \mathrm{arg min}_U \,\eta(x,\Sol{p}{u}(\Delta t))$ wrt.  $O$.
\item Let $p_\mathrm{new} = p + \Sol{p}{u}(\Delta t)$.
\item If $\eta(q,p_\mathrm{new}) < \eta(q,p)$ then
\begin{itemize}
\item[] Add $p_\mathrm{new}$ to $\mathcal{R}$ with a connection from $p$ \hspace*{2cm}
\item[] Update $p$ to $p_{\mathrm{new}}$
\end{itemize}
else
\begin{itemize}
\item[] Exit the loop
\end{itemize}
\end{enumerate}
\end{algorithm}

\section{Computational Experiments}
\label{sec:exp}

We implemented the exploring LQR-tree algorithm in MATLAB 2020b on a~PC with Intel Core i7-10700K with 32GB of RAM. We constructed our NLP demonstrator using the toolbox CasADi~\cite{Andersson2018}  with the internal optimization solver Ipopt~\cite{ipopt}, performing discretization via the Hermite-Simpson collocation method~\cite{Har:87}. 

The implementation uses a slightly modified assignment rule $\phi_\mathrm{LQR}$ that reduces its computation time by first narrowing the search of the target demonstration to the $N = 500$ closest demonstrations according to the Euclidean distance.

In Algorithm~\ref{LQRRRT} we terminate the loop in step 2 whenever either a new demonstration is discovered (both in the counterexample tree and the demonstration tree), or a maximum number of extensions of any given tree is exceeded. 

In RRT algorithms, the choice of a good metric $\eta$ for selecting which state to expand is heavily problem dependent~\cite{RRT5}. In our computational experiments, we used the weighted Euclidean metric augmented with the history-based selection~\cite{RRT17}. In this modification, the number of failed extensions of any given state is tracked. A new history based metric $\eta^H$ is defined as
\begin{equation}
\eta^H(x,x_\mathrm{rand}) = \frac{\eta(x,x_\mathrm{rand}) - \eta_\mathrm{min}}{\eta_\mathrm{max} - \eta_\mathrm{min}} + \frac{n-n_\mathrm{min}}{n_\mathrm{max}-n_\mathrm{min}},
\end{equation}
where $n$ is a number of failed extensions of $x$, $n_\mathrm{max}$ is a maximum number of failed extensions, $n_\mathrm{min}$ is a minimum number of failed extensions, $\eta_\mathrm{max}$ is the maximum distance between $x_\mathrm{rand}$ and nodes of the tree, and  $\eta_\mathrm{min}$ is the minimum distance between $x_\mathrm{rand}$ and nodes of the tree. 

Our implementation also modifies Algorithm~\ref{LQR-tree gen} to not only terminate  when a new demonstration is successfully generated from $x_\mathrm{counter}$, but also when the counter-example tree reaches a  certain maximal number of nodes.   If this happens, Algorithm~\ref{LQR-tree gen} is restarted from the next detected counterexample. This avoids excessive growth of the RRT-trees and a resulting slow-down due to costly nearest neighbour computations.

All demonstrations require a given clearance.  However, this may make the generation of a new demonstrations difficult for the LQR-tree feedback control law, since it is based on the tracking of some previous demonstration which could meet the clearance condition only tightly. Hence, we added a certain tolerance (5\% in our implementation) for obstacles and state bounds that is sufficient for the Algorithm~\ref{LQR-tree gen} to pass the solution to $demo$. Naturally, the NLP demonstrator must produce the new demonstration with the clearance as prescribed, which always happened in our experiments.

We consider an universal time step $h$, which we use both in computations of demonstrations for time discretization and in the RRT algorithm as a simulation step. Finally, in all numerical experiments, we terminated the LQR-tree algorithm after reaching 1000 samples without a new counterexample.

\subsection{Examples}
In our numerical experiments, we considered the following problems.

\paragraph*{Inverted pendulum} We return to the inverted pendulum example~\eqref{inverted_pendulum}. We restrict the demonstrations by state bounds $X'_\mathrm{B} = \pm\left[8,12\right]$. We set the time step $h$ as $0.05$ and the length of demonstrations to $T = 10$. In the RRT algorithm, we used the standard Euclidean distance without history based sampling and we reduced the control inputs to the two values $-1$ and $1$. We set the maximum number of extensions of a single tree within a single bidirectional expansion, that is, before a new evaluation of the closest demonstration, to $500$ and the maximum size of a counterexample tree to $5000$.

\paragraph*{Inverted pendulum on a~cart} We use a slightly modified existing example~\cite{LQRtrees2}. Here, we order the variables as $\left[ x, \theta, \dot{x}, \dot{\theta}\right]$. We use the initial set $I = \left[-0.2, 0.2\right] \times \left[-\pi, \pi\right] \times \left[-1.5, 1.5\right] \times \left[-8, 8\right]$, state bounds $X_\mathrm{B}\pm\left[0.45,4\pi,15,50\right]$, control bounds $U = [-60, 60]$, and goal set $G = \{x\in\mathbb{R}^4 \mid x^T\mathrm{diag}(10,1,1,1)x < 0.05\}$. We set the length of demonstrations to $T = 7.5$ and restrict them to state bounds $X'_\mathrm{B} = \pm\left[0.36,2\pi,8,25\right]$ and the control inputs to $U' = [-36, 36]$.

The cost matrices of the LQR tracking $Q = \mathrm{diag}(100,30,100,10)$ and $R = 0.1$, and use the same cost for generation of demonstrations in the NLP demonstrator. We set the time step $h$ to $0.025$. In Algorithm~\ref{LQRRRT}, we used the standard Euclidean distance with the history-based selection~\cite{RRT17} and we reduced the input selection to the two values $-36$ and $36$. We set the maximum number of extensions of a single tree within a single bidirectional RRT expansion to $500$ and maximum size of a counterexample tree to $5000$.

\paragraph*{Quadrotor} We use the simplified model from~\cite[(2.61)--(2.66)]{quad_model}, where we set $m = 1$ and $g = 9.8$ and we order the state variables as $\left[x,y,z,\phi,\psi,\theta,\dot{x},\dot{y},\dot{z},\dot{\phi},\dot{\psi},\dot{\theta}\right]$. We bound the state variables $x$ and $y$ by $-\left[8,12\right] \times \left[12,12\right]$ and set bounds for the rest to $\pm\left[5,1,4\pi,1,10,\ldots,10\right]$. We bound the control inputs by $U = [-2.5, 2.5]^4$. We also consider three cylindric
obstacles with bases $\left\{y\in\mathbb{R}^2\mid (y-a)^TA(y-a)\leq \alpha\right\}$, where we set $a = [2,-3.5],[2,3.5],[2,0]$, $\alpha = 8,8,7$ and $A = \mathrm{diag}(1,45),\mathrm{diag}(1,45),\mathrm{diag}(25,1).$ 

\begin{figure}
\centering
\begin{subfigure}{0.45\textwidth}
\centering
\includegraphics[width=1\textwidth]{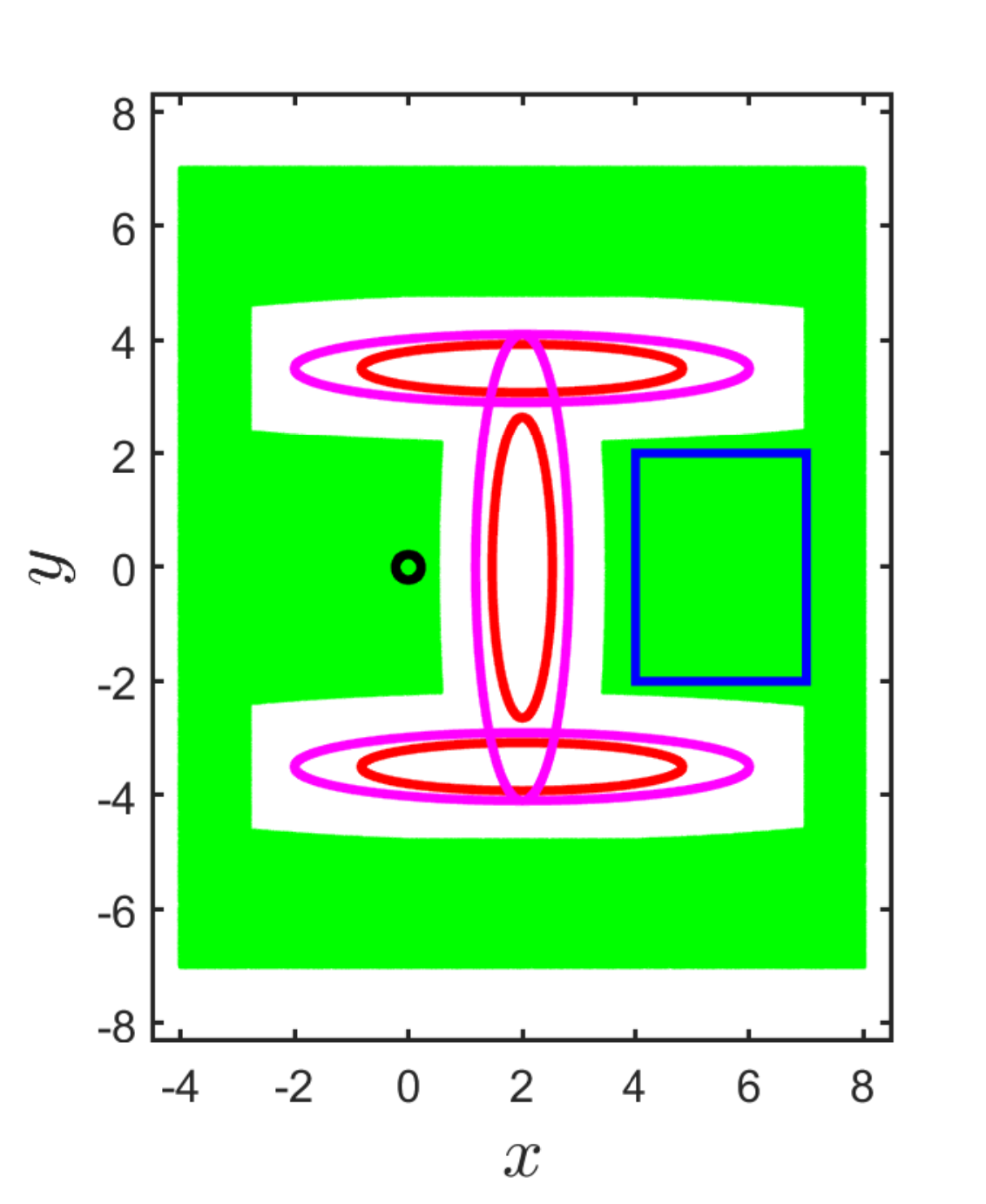}
\end{subfigure}
\centering
\begin{subfigure}{0.45\textwidth}
\includegraphics[width=1\textwidth]{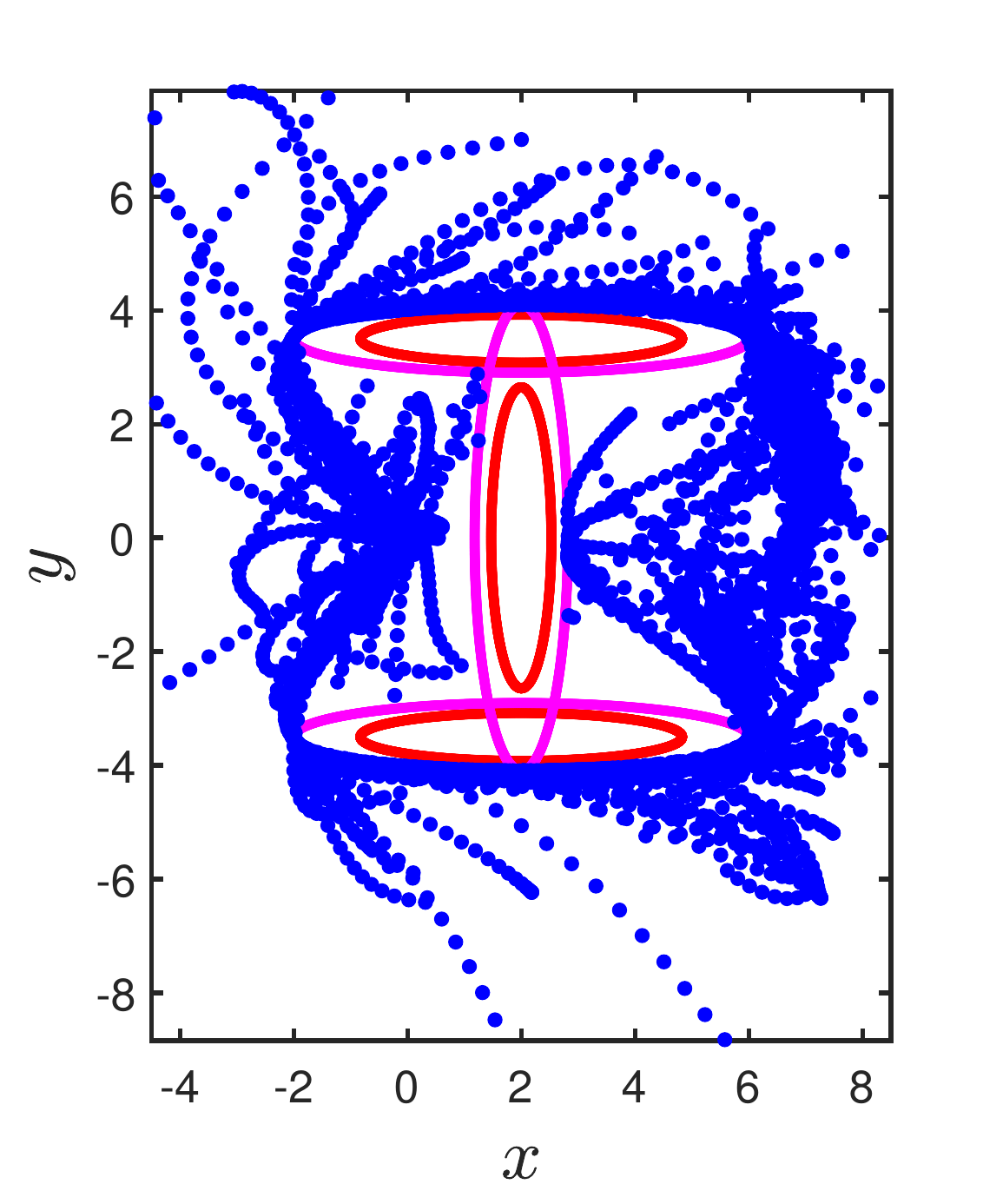}
\end{subfigure}
\caption{Quadrotor example: Initial set 1 (blue rectangle), initial set 2  (green area), obstacles (red, magenta for demonstrations) and the goal set (black circle).  The final LQR-tree produced by the exploring LQR-tree algorithm for the initial set 2.}
\label{quad_ex}
\end{figure}

We consider two variants of this problem with different initial sets. The first variant uses a rectangle $[4,7]\times [-2,2]$, and the second one enlarges the initial set to subset of $[-4,8] \times [-7,7]$, see Fig.~\ref{quad_ex}. In both variants all the other variables are set to zero. We use the goal set $G = \{x\in\mathbb{R}^{12} \mid \norm{x} < 0.05\}$. We set length of demonstrations to $T = 10$ and restrict them to $-\left[5,9\right] \times \left[9,9\right]$ in $x$ and $y$. The bounds for the rest are $\pm\left[0.5,2\pi,0.5,5,5,5,2,2,2\right]$. The other values are restricted to $U' = [-2, 2]^4$ and  $\alpha' = 16,16,16.8$. 

The cost matrices of LQR tracking are $Q = I$ and $R = I$. The same cost was used for generation of demonstrations in the NLP demonstrator. We set the time step $h$ to $0.1$. In the RRT algorithm~\ref{LQRRRT}, we used the weighted Euclidean distance with $W = \mathrm{diag}(5,5,1,\ldots,1)$ with the history-based selection~\cite{RRT17} and we reduced  each input selection to the two values $-2$ and $2$ considering all its combinations. We set the maximum number of extensions of a single tree within a single bidirectional RRT expansion to $500$ and maximum size of a counterexample tree to $5000$.

\subsection{Results}

\begin{table*}
\center
\begin{tabular}{l  || r r r || r r r || r r r}
&\multicolumn{3}{c||}{exploring LQR-tree}&\multicolumn{3}{c||}{simple LQR-tree}&\multicolumn{3}{c}{funnel LQR-tree}\\
\hline
Ex.  & $N_\mathrm{dem}$ & $s_\mathrm{dem}$ &$t_{\mathrm{total}}$ & $N_\mathrm{dem}$ & $s_\mathrm{dem}$ &$t_{\mathrm{total}}$ & $N_\mathrm{dem}$ & $s_\mathrm{dem}$ &$t_{\mathrm{total}}$\\
\hline	
Pnd   & 36 & 100\% & 1.1 &  39 & 4\% & 10.6 &  18 & 1\% & 17.2\\
Pndc & 557  & 100\% & 27.4 &  $^*$1365 & $^*$50\% & TO & $^*$224 & $^*$34\% & $^*$548.1\\
Quad\_initial\_set1 & 139  & 100\% & 62.9 & $^*$0  & $^*$0\% & TO & $^*$0  & $^*$0\% & TO \\
Quad\_initial\_set2 & 276  & 100\% & 114.5 & 12 & 10\% & 46.4 & 14 & 3\% & 196.7\\
\hline
\end{tabular}
\caption{Number of added demonstrations $N_\mathrm{dem}$, success rate of the NLP demonstrator $s_\mathrm{dem}$ and total CPU time in minutes $t_{\mathrm{total}}$. A star denotes that some runs exceed 10 hours CPU time mark (TO). All values are averages using 5 runs.}
\label{tab:res}
\end{table*}

We compare the exploring LQR-tree algorithm with two reference algorithms. In the first reference algorithm, which we call the \emph{simple LQR-tree algorithm}, we use a simple heuristic, initializing the demonstrator with the system simulation that detected the given counterexample $x$. We also implemented an LQR-tree algorithm as described in~\cite{LQRtrees2} which we will further call  the \emph{funnel LQR-tree algorithm}. This algorithm uses a slightly different assignment rule to ours
\begin{equation}
\label{LQRarulemod}
\phi(x,\mathcal{T}) =\underset{{(\tilde{x},\tilde{u})\in\mathcal{T}}}{\mathrm{arg\,min}} \left\{V_{(\tilde{x},\tilde{u})}(x)\,\text{s.t.}\, V_{(\tilde{x},\tilde{u})}(x) \leq d_{(\tilde{x},\tilde{u})}\right\},
\end{equation}
where the values $d_{(\tilde{x},\tilde{u})}$ represent estimations of so called funnels~\cite{LQRtrees2} and are computed as follows: when the algorithm detects a counterexample, it lowers the value of the corresponding $d_{(\tilde{x},\tilde{u})}$ such that the counterexample is no longer assigned to the demonstration. Then the algorithm tries to assign a new demonstration via~\eqref{LQRarulemod}. Provided that no simulation is successful, it chooses one failed simulation based on the cost used in the NLP demonstrator. If some state is not assigned to any demonstration, the algorithm proceeds the same way as the simple LQR-tree algorithm.  

For all three algorithms, all results are averages over 5 runs. As can be seen from these results, initiating demonstrations from failed system simulations (the simple LQR-tree and the funnel LQR-tree algorithms) can lead to poor performance of the NLP demonstrator, as was already observed in~\cite{LQRtrees2}. This can result in an excessively large runtime of the overall algorithm and even failure of the construction of the LQR-tree itself, when the heuristic is unable to make progress. In comparison, the demonstrator in the exploring LQR tree algorithm did not fail even once and all problems were solved. In the Example 3, the simple LQR tree algorithm actually performed better overall, but for the cost of enlarging the initial set to ensure that the heuristic could make progress.

The main disadvantage of the exploring LQR-tree algorithm is that it may generate a significantly larger LQR-trees. The reason for this is that the exploring LQR-tree algorithm can add new demonstrations even outside of the initial set. This allows the growth of the LQR-tree to be more gradual, but computing these additional demonstration can be costly (in Example 3, it was over 50\% of the overall runtime). Hence, it may be beneficial to add further restrictions on adding demonstrations outside of the initial set. The resulting LQR-trees can also be pruned from these additional demonstrations once the initial set is covered.

\section{Conclusion}
\label{sec:conclusion}
In this paper, we presented the exploring LQR-tree algorithm that incorporates a bidirectional RRT algorithm for generation of initial estimates of demonstrations. These estimate are subsequently handled by the demonstrator. In comparison to the previous work, which included a simple heuristics that used failed simulations, this approach is more reliable, often faster and allows solution to problems that could not be handled by the previous algorithms.

\bibliographystyle{plain}
\bibliography{lqr_trees_sampling}

\end{document}